\documentclass[useAMS,usenatbib]{mn2e}
\usepackage{psfig,url,morefloats}
\usepackage{subfig}
\usepackage{graphicx}
\usepackage{float}
\usepackage{gensymb}

\footnotesize
\newdimen\digitwidth    
\setbox0=\hbox{\rm0}
\digitwidth=\wd0
\catcode`!=\active
\def!{\kern\digitwidth}
\normalsize

\title[Simultaneous radio and X-ray observations of PSR~B0611+22]
  {Simultaneous radio and X-ray observations of PSR~B0611+22}
\author[K. Rajwade et al.]
  {K.~Rajwade,$^1$\thanks{email: kmrajwade@mix.wvu.edu}
  A.~Seymour,$^2$ D.R.~Lorimer,$^1$ A.~Karastergiou,$^3$ M.~Serylak,$^{3,4,5}$
  \newauthor
   M.A.~McLaughlin$^1$ and
   J-M.~Griessmeier$^{4,5}$
  \\
  $^1$Department of Physics and Astronomy, West Virginia University, Morgantown, WV 26506,USA\\
  $^2$National Astronomy and Ionospheric Center, Arecibo Observatory, Arecibo, PR 00612, USA\\
  $^3$Astrophysics,University of Oxford,Denys Wilkinson Building, Keble Road, Oxford OX1 3RH, UK\\
  $^4$Department of Physics \& Astronomy, University of the Western Cape, Private Bag X17, Belville 7535, South Africa \\
  $^5$LPC2E - Universit\'{e} d'Orl\'{e}ans / CNRS, 3A, Avenue de la Recherche Scientifique, 45071 Orléans cedex 2, France\\
  $^6$Station de Radioastronomie de Nan\c{c}ay, Observatoire de Paris - CNRS/INSU, USR 704 - Univ. Orl\'{e}ans,\\
OSUC, route de Souesmes, Nan\c{c}ay, France}
\date{Released 2002 Xxxxx XX}

\pagerange{\pageref{firstpage}--\pageref{lastpage}} \pubyear{2002}

\def\LaTeX{L\kern-.36em\raise.3ex\hbox{a}\kern-.15em
    T\kern-.1667em\lower.7ex\hbox{E}\kern-.125emX}

\begin{document}

\label{firstpage}

\maketitle

\begin{abstract}
 We report results from simultaneous radio and X-ray observations of 
 PSR B0611+22 which is known to exhibit bursting in its single-pulse
 emission. The pulse phase of the bursts vary with radio 
frequency. The bursts are correlated in 327/150~MHz datasets while 
they are anti-correlated, with bursts at one frequency associated with normal 
emission at the other, in 820/150~MHz datasets. Also, the flux density of this pulsar is lower than expected at 327~MHz assuming a power law. We attribute this unusual behaviour
to the pulsar itself rather than absorption by external astrophysical sources. 
Using this dataset over an extensive frequency
 range, we show that the bursting phenomenon in this pulsar exhibits temporal variance over a span of few hours. 
We also show that the bursting is quasi-periodic over the observed band. The anti-correlation in the phase offset of the burst mode at different frequencies suggests that the mechanisms 
responsible for phase offset and flux enhancement have different dependencies on the frequency. We did not detect
 the pulsar with \textit{XMM-Newton} and place a 99$\%$ confidence upper limit on the
 X-ray efficiency of $ 10^{-5}$. 
\end{abstract}

\section{Introduction}

The 0.33~s pulsar PSR B0611+22 (characteristic age $\sim$90 kyr) was discovered by \cite{Da72} and was 
initially thought to be associated with the supernova remnant (SNR) 
IC~443 which lies at close angular separation to the pulsar~\citep{Da72, Hi72}. 
This association was always doubtful as the pulsar lies well beyond the radio shell \citep{Du75} of the remnant. Recent X-ray observations detected a compact 
X-ray source within the remnant shell and the corresponding pulsar wind 
nebula \citep{Ch01} which rejected any association of the pulsar with the remnant. 
Moreover, IC~443 is known to lie within the molecular cloud G189$+$3.3~\citep{Bo00} which lies along the line of sight to the pulsar. Although, the distances to these sources are highly uncertain, it is reasonable to assume that the pulsar lies beyond these dense regions~\citep{Fe84,We03}. This suggests that 
the radio emission propagates through the dense medium which might contribute to the pulsar's dispersion measure (DM) of $\sim$96~${\rm pc}~{\rm cm^{-3}}$. The environment of this pulsar makes it an 
interesting object for studies of radio emission and single-pulse properties.

The pulsar was studied by \cite{No92}, who found that PSR B0611+22 appears to exhibit different modes in which the enhanced emission mode peaked at a later pulse phase than the average profile and the weak mode peaked at an earlier phase.
Recently, \cite{Se14} performed a detailed study of the
emission behaviour of PSR~B0611+22. 
They found that, at 327~MHz, the pulsar shows
steady emission in one mode which is enhanced by bursting emission that is
slightly offset in pulse phase from this steady
emission. \cite{Se14} also observed the bursting to be quasi-periodic with a period around $\sim1000$ pulse periods. This type of
behaviour has also been seen in other pulsars like PSR~J1752+2358~\citep{Ga14}
and PSR~J1938+2213 \citep{Lo13}. PSR B0611+22's short mode changes with offset in the emission
phase could be responsible for the high degree of timing noise the pulsar exhibits~\citep{Ar94}. 

The phenomena of nulling and mode changing which relate to such
emission behaviour have been studied in different pulsars for four
decades. They were first observed and reported by Backer \citep{Ba70d, Ba70c, Ba70b, Ba70a}. Mode changing pulsars are pulsars in which,
from time to time, the mean profile abruptly changes between two or more quasi-stable states
\citep{Wa07, Ba82} while nulling is the
abrupt cessation of radio emission for one or more pulse periods. Nulling has
been postulated to be an extreme case of mode changing \citep{Wa07,Ti10}. In a series of papers, 
Rankin \citep{Ra83,Ra86,Ra03} tried to understand the emission
geometry and behaviour of such pulsars. According to her model, the emission beam 
of a pulsar consists of a central core emission beam surrounded by 
multiple annular cones of emission. The pulse profile we observe depends on 
which core and/or cone beams are traversed by the line of sight of the observer. Rankin suggested that mode
changing can be thought of as a reorganization of such core and conal
emission resulting in a change in the observed pulse profile.
Mode changing has been observed in most multi-component pulsars
(pulsars with more than one component in their emission profile)~\citep{Ra86}. Many
pulsars like PSR B2319+60 \citep{Wr81}, PSR B0943+10 \citep{Su98} and
PSR B1918+19 \citep{Ra13} exhibit this
phenomenon. Both nulling and mode changing have been studied in $\sim$200 pulsars so far \citep{Bi92, We06, Wa07,Ga12}. PSR~B0611+22 has been classified as a core
emission pulsar with a single component~\citep{Ra83}. This makes the pulsar interesting as the phase offsets and flux enhancement are small 
in comparison to other pulsars in terms of magnitude and timescale and are 
harder to explain in the standard framework.
Recently, a global picture of quasi-stable states of the magnetosphere has come to the fore~\citep{Ly10,He13}.
\cite{He13} discovered an anti-correlation between X-ray and radio 
emission in the two modes of emission of PSR~B0943+10. This result 
motivated us to ask whether such X-ray emission is also detectable in 
PSR~B0611+22 and, if yes, how does it relate to the mode changes seen in radio? This led to a simultaneous radio and X-ray observation campaign of PSR~B0611+22.

As mentioned above, PSR B0611+22 has a supernova remnant and a molecular cloud in its vicinity. Such dense environments
around and likely, in front of the pulsar make it an ideal candidate to study the effects of these environments on the measured flux density. Previously, pulsars within such
dense environments have been known to show a spectral turnover at 
frequencies around $\sim1$~GHz~\citep{Ki07, Ki11}. A recent study by~\cite{Ra15} shows that it is possible to derive the physical parameters of these dense regions by modeling the flux density spectrum of the pulsar. In this paper, we try to characterize the peculiar
emission behaviour with a multi-wavelength, broadband dataset of the pulsar. 
The observational details are given in Section
2. The results are presented in Section 3. The discussions based on the results are in Section 4. The conclusions are given in Section 5.

\section{Observations}

All observations were carried out on MJD 56756. PSR~B0611+22
was observed at three different radio frequencies including 327~MHz (Arecibo
Observatory), 820~MHz (Green Bank Telescope) and 150~MHz (International LOFAR station-Nan\c{c}ay, France). The observation configurations for the radio
telescopes are given in Table~\ref{tab:config}. The data were recorded and converted into multi-channel
filterbank format before being written out to disk. Then, the data were
incoherently dedispersed using \textbf{PRESTO}\footnote{\url{http://www.cv.nrao.edu/~sransom/presto}} at the pulsar DM of 96 ${\rm pc}~{\rm cm^{-3}}$ to remove the dispersion delay of
incoming radio waves due to the interstellar medium. For the LOFAR (FR606) data, coherent dedispersion was carried out using \textbf{DSPSR}\footnote{\url{http://dspsr.sourceforge.net}}~\citep{Va11}.
\\
\begin{table*}
\begin{tabular}[ht]{lc lc lc lc lc lc}
\hline
Telescope & $\nu$ & $t_{\rm samp}$ & $\Delta\nu$ & $t_{\rm int}$ & $n_{\rm chan}$ & $G$ & $T_{\rm sys}$ & UT \\
          &  (MHz)    &          (ms) & (MHz)     & (hrs)&  & (K/Jy) & (K) & hh:mm \\
\hline
Green Bank Telescope (GBT) & 820 & 0.15 & 200 & 1.25&2048 &2&101&16:30 - 17:45\\
Arecibo Observatory (AO) & 327 & 0.5 & 50 & 2&2048 &11&117& 20:30 - 22:30 \\
LOFAR (FR606) & 148 & 0.32& 80 & 6&400&0.97& 900& 16:45 - 22:30 \\
XMM-Newton & & & & & & & & 16:30 - 22:30 \\
\hline
\end{tabular}
\caption{Configuration of different radio telescopes during observation of PSR B0611+22. From left to right, we list the observing frequency ($\nu$), sampling time ($t_{\rm samp}$), bandwidth ($\Delta\nu$), observation time ($t_{\rm int}$), number of frequency channels ($n_{\rm chan}$) gain ($G$), system temperature ($T_{\rm sys}$) and Universal Time (UT) of observation.
\label{tab:config}}
\end{table*}
\\


Simultaneously, the pulsar was observed by {\it XMM-Newton}.
The {\it XMM-Newton} observations 
used the photon imaging camera (EPIC)~\citep{St01,Tu01}.
The PN-CCD was operated in
small-window mode with a medium filter to block stray optical
light. All the events recorded by the PN camera are time-tagged with a
temporal resolution of 5.7~ms. On the other hand, the MOS CCDs were operated in full-window mode with a medium filter in
each camera, which provide us with a large field-of-view. The coverage of various telescopes during the whole observation is illustrated in
Table.~\ref{tab:config}.

\begin{figure*}
\centering
\centerline{\psfig{file=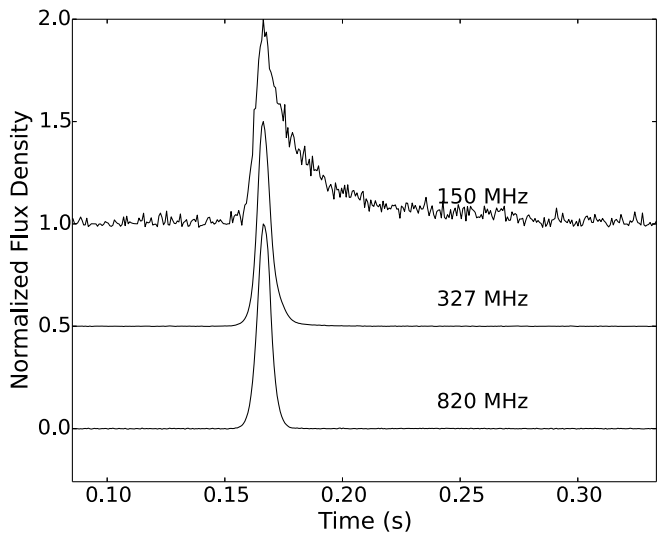,width=100mm}}
\caption{The peak of the average profile of PSR B0611+22 at different radio frequencies. For clarity, only part of the period near the peak is plotted.}
\label{fig:Avg_prof}
\end{figure*}


\section{Results}
\subsection{Broadband bursting} 

We analyzed the radio data for bursting behaviour at different
frequencies similar to~\cite{Se14}. For each frequency, namely 820, 327 and 150~MHz, 
the time series were folded at the topocentric period of the
pulsar to generate the averaged pulse profiles shown in Fig.~\ref{fig:Avg_prof}. 

\begin{figure*}
\centering

\centerline{\psfig{file=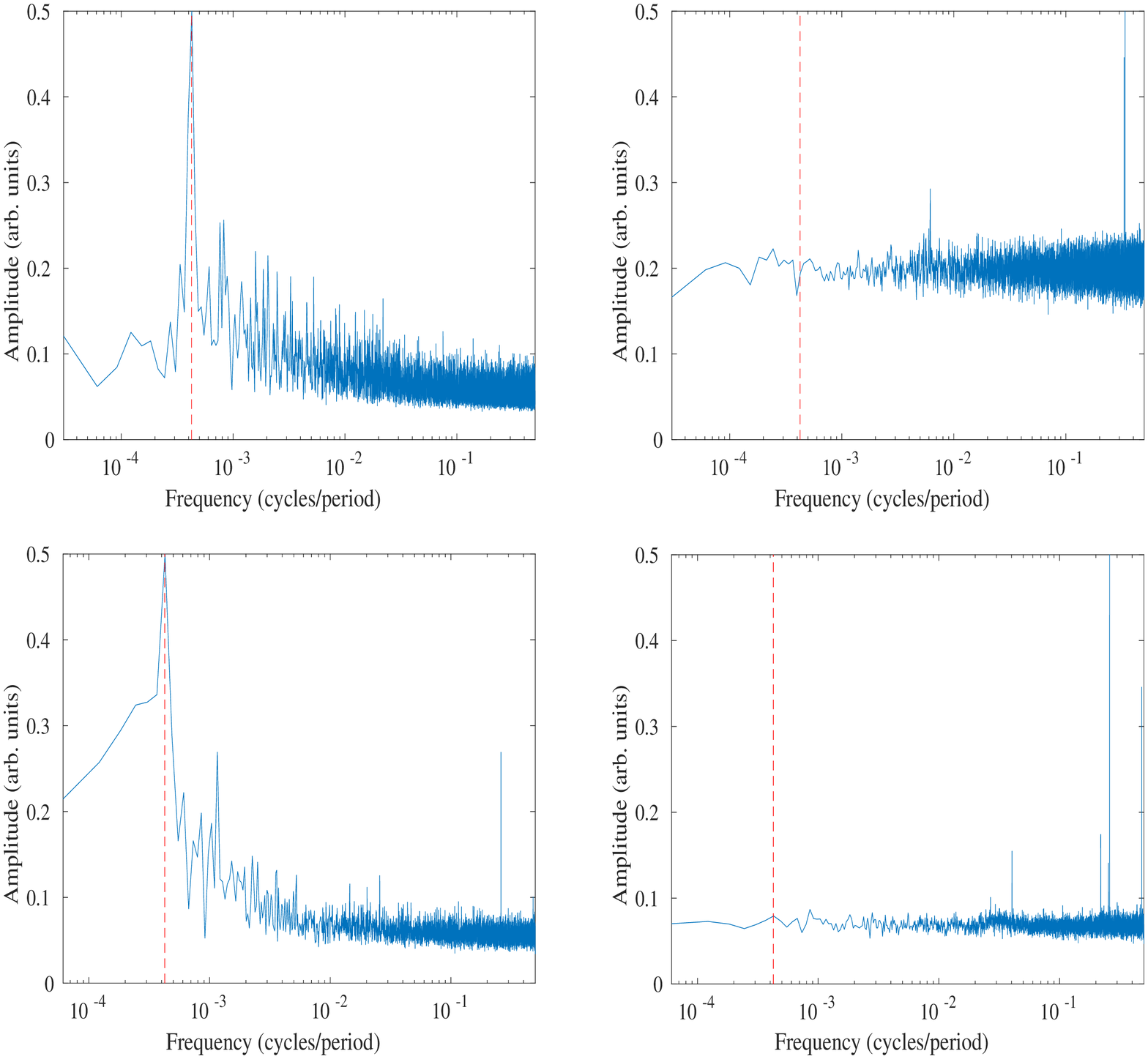,width=150mm}}


\caption{Fluctuation spectrum for PSR B0611+22 for \textbf{Top Panels}: 327 MHz and 
\textbf{Bottom Panels}: 820 MHz. The left panels show the integrated On pulse power 
spectrum and the right panels show the integrated Off pulse power spectrum. 
The red vertical line corresponds to a frequency of $\sim$0.0004 cycles/period. One can see the spike at $\sim$0.0004 cycles/period in the top panels corresponding to a period of 2500 pulse periods at both frequencies.
\label{fig:flucspec}}
\end{figure*}

\begin{figure*}
\centering
\begin{tabular}{|@{}l|@{}l|r@{}|}

{\mbox{\psfig{file=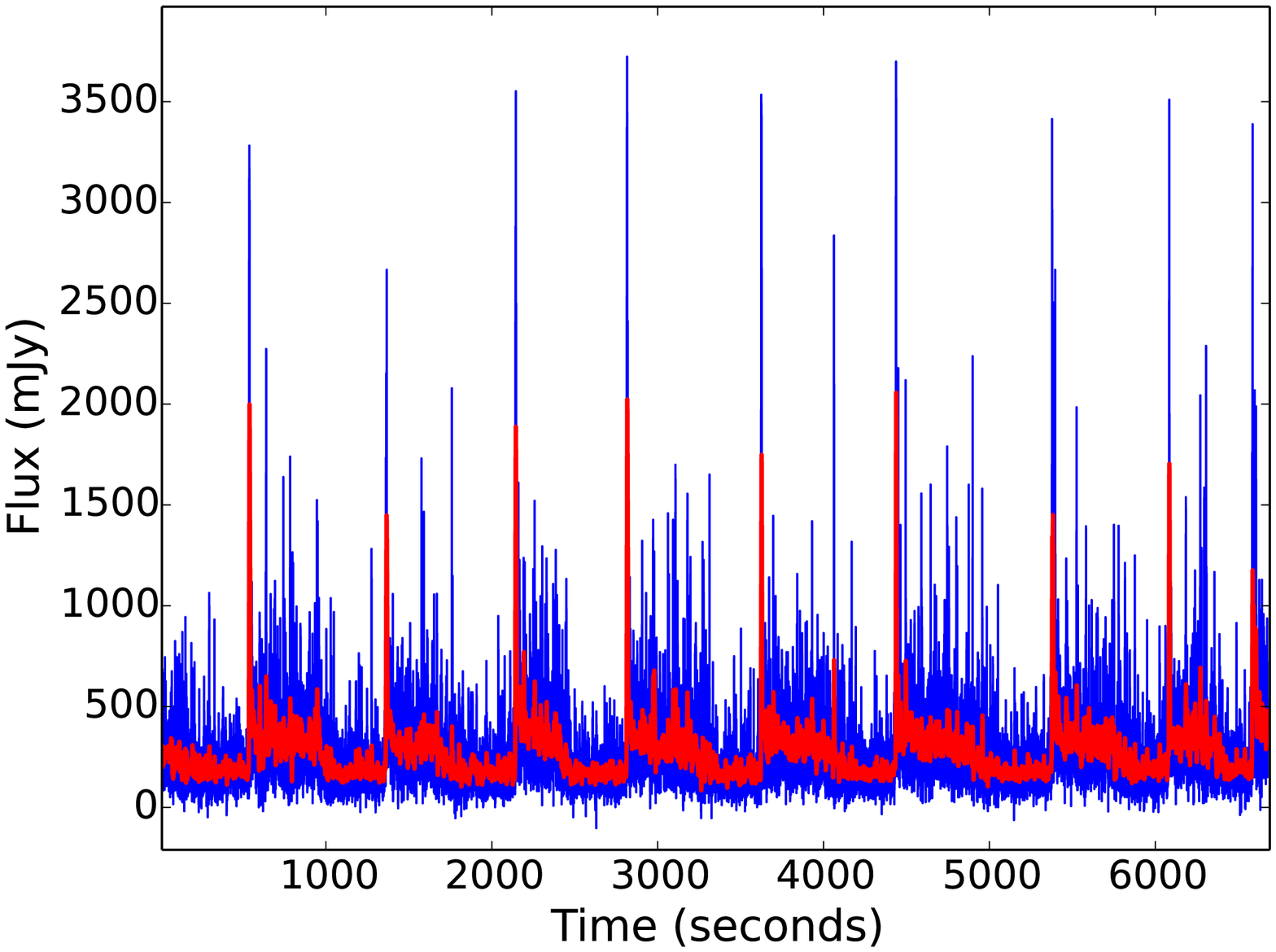,width=100mm,height=80mm}}} &{\mbox{\psfig{file=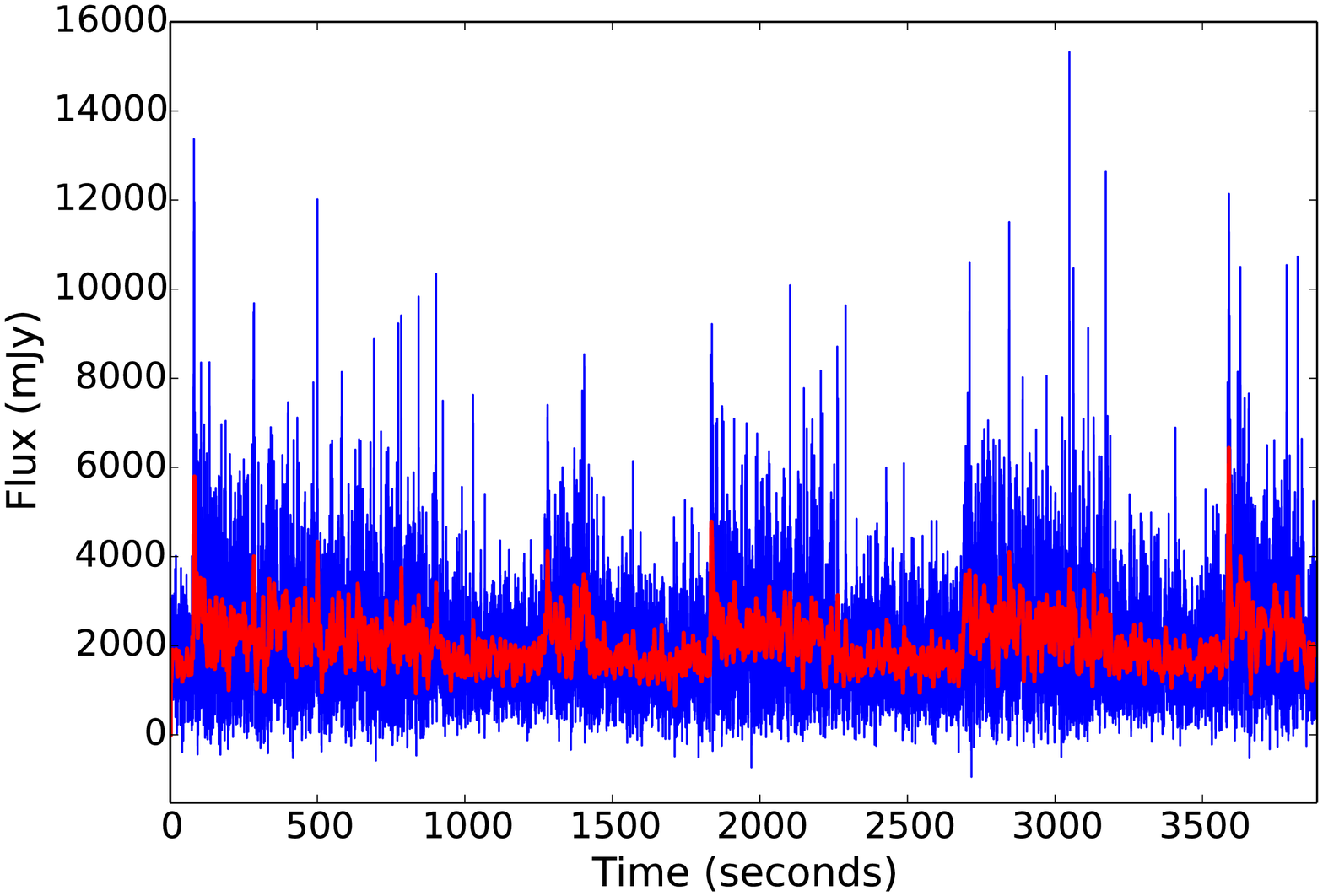,width=100mm,height=80mm}}}
\\

\end{tabular}

\caption{Flux time series of a single bin of PSR~B0611+22~\textbf{Left Panel}: 327 MHz (duration $\simeq$2 hours) and~\textbf{Right Panel}: 820 MHz (duration $\simeq$1.25 hours). The blue line is the full resolution data. The red line is the smoothed version.
\label{fig:Ener_1bin}}
\end{figure*}

\begin{figure*}
\centering
\begin{tabular}{|@{}l|@{}l|r@{}|}

{\mbox{\psfig{file=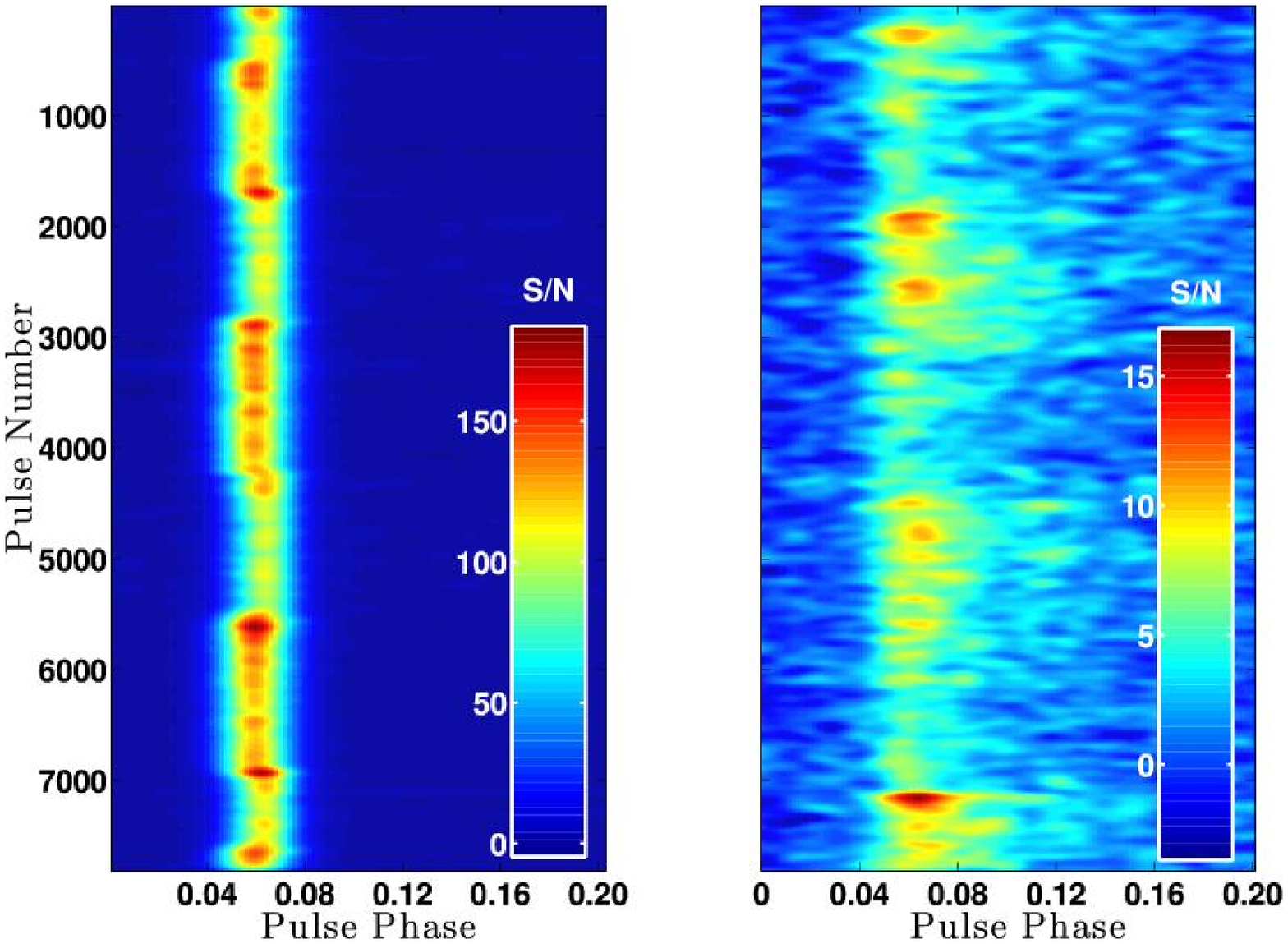,width=80mm}}} &{\mbox{\psfig{file=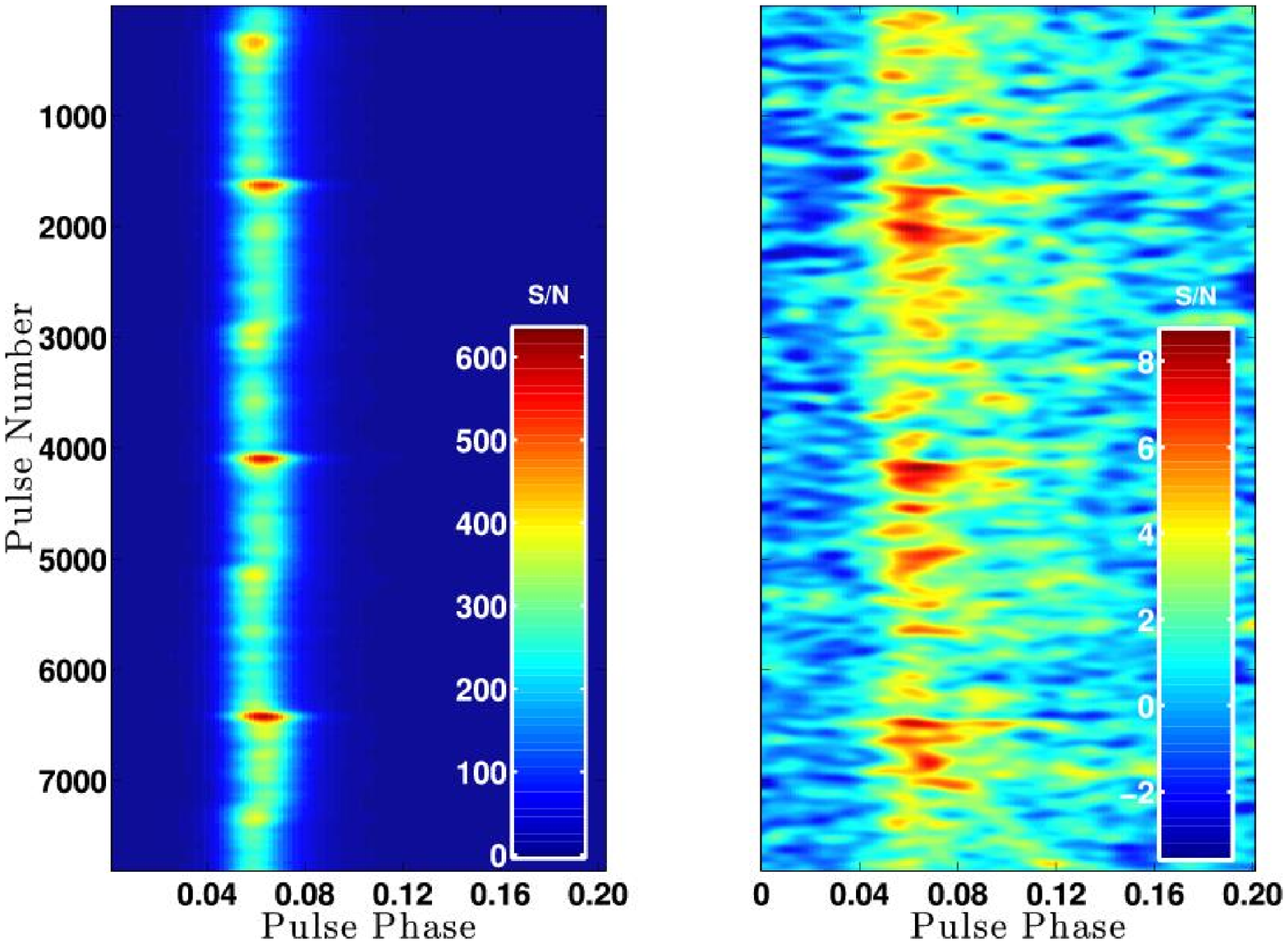,width=80mm}}}
\\

\end{tabular}

\caption{Simultaneous radio observations of PSR~B0611+22 showing 820/150~MHz observations (left) and
327/150~MHz observations (right). The negative S/N arises due to fluctuations of off-pulse noise below zero mean. Both datasets are of the same duration. Each dataset has been smoothed by using a 2-D Gaussian kernel of same
dimensions (see text for details).}
\label{fig:LOFAR_obs}
\end{figure*}

For these folded time series (see Fig.~\ref{fig:LOFAR_obs}), we Fourier transformed
the energies at each pulse longitude (phase bin) to obtain a longitude resolved
fluctuation spectrum for 327 and 820 MHz. The fluctuation spectrum was integrated over specific On and Off pulse windows to obtain the integrated power spectrum for both regions as shown in Fig.~\ref{fig:flucspec}. The figure suggests that bursts seem to be quasi-periodic at both frequencies. We derived a rough periodicity for the
bursts of $\sim$2500 pulse periods from Fig.~\ref{fig:flucspec}. The periodic nature of emission is evident in Fig.~\ref{fig:Ener_1bin}
which shows how energy of a single bin corresponding to the peak in the average profile varies in time. The profile was chosen
from the ON pulse window at 327 and 820 MHz as done in~\cite{Se14}.

Due to telescope scheduling constraints, there was no overlap between
the 327 and 820~MHz observations. However, each of those observations
overlapped with the 150~MHz observations so we decided to compare the observations at 327 MHz and 820~MHz with the corresponding spans in the 150~MHz
observations. Since we do not have high enough sensitivity from the
LOFAR (FR606) observations to detect single pulses, we decided to convolve the pulse stack using a 2-D Gaussian kernel with a 
width spanning 64 time bins (i.e. single pulses) and 5 pulse phase bins. We
applied the same smoothing to each dataset to give them comparable resolution. 
To ensure time alignment, we removed dispersion delay for each telescope using 
infinite frequency as the reference. The timeseries for each dataset were barycentered to remove any time delays due to different locations of the telescopes.

\begin{figure*}
\centering

\centerline{\psfig{file=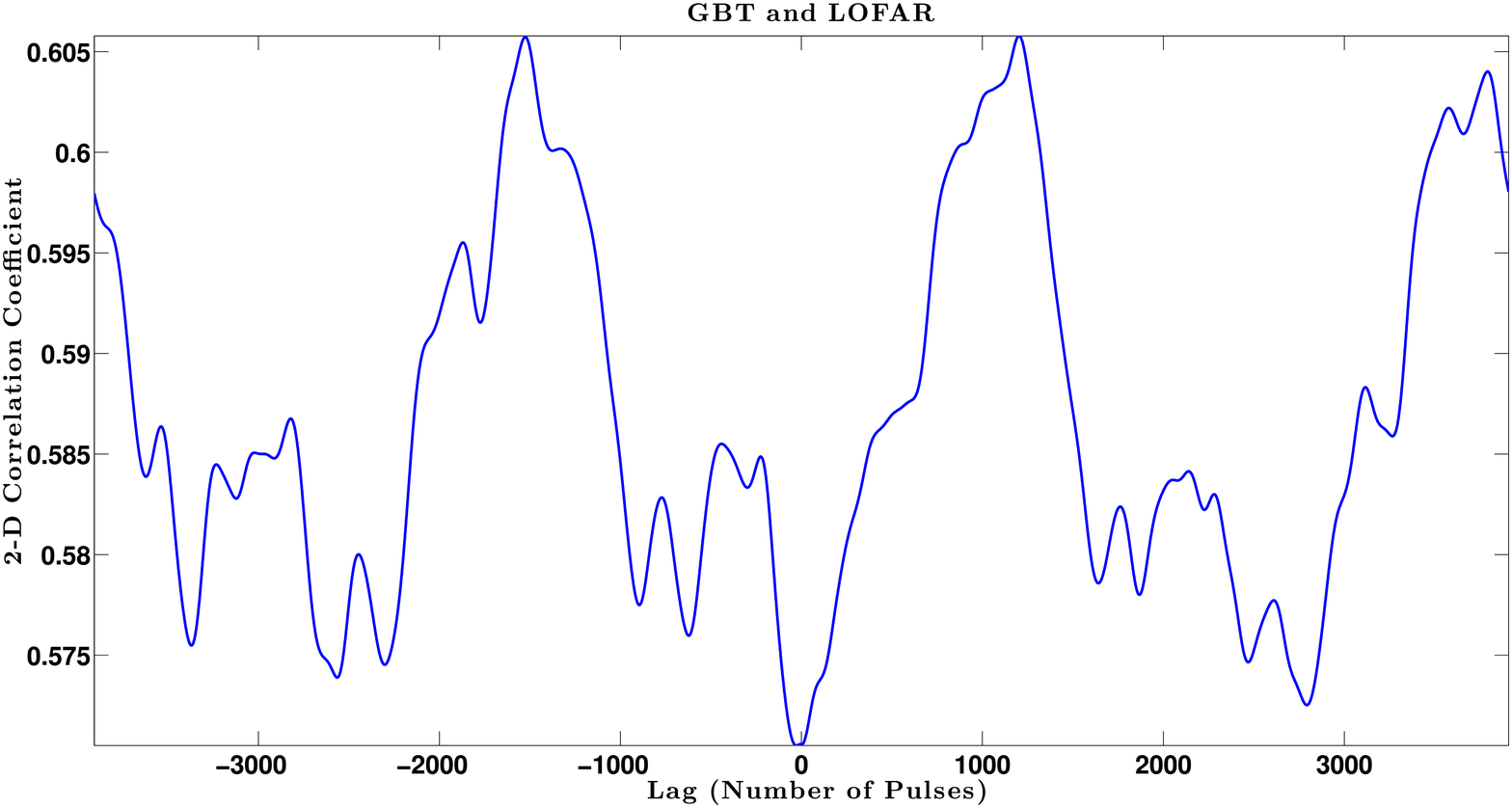,width=150mm}}
\centerline{\psfig{file=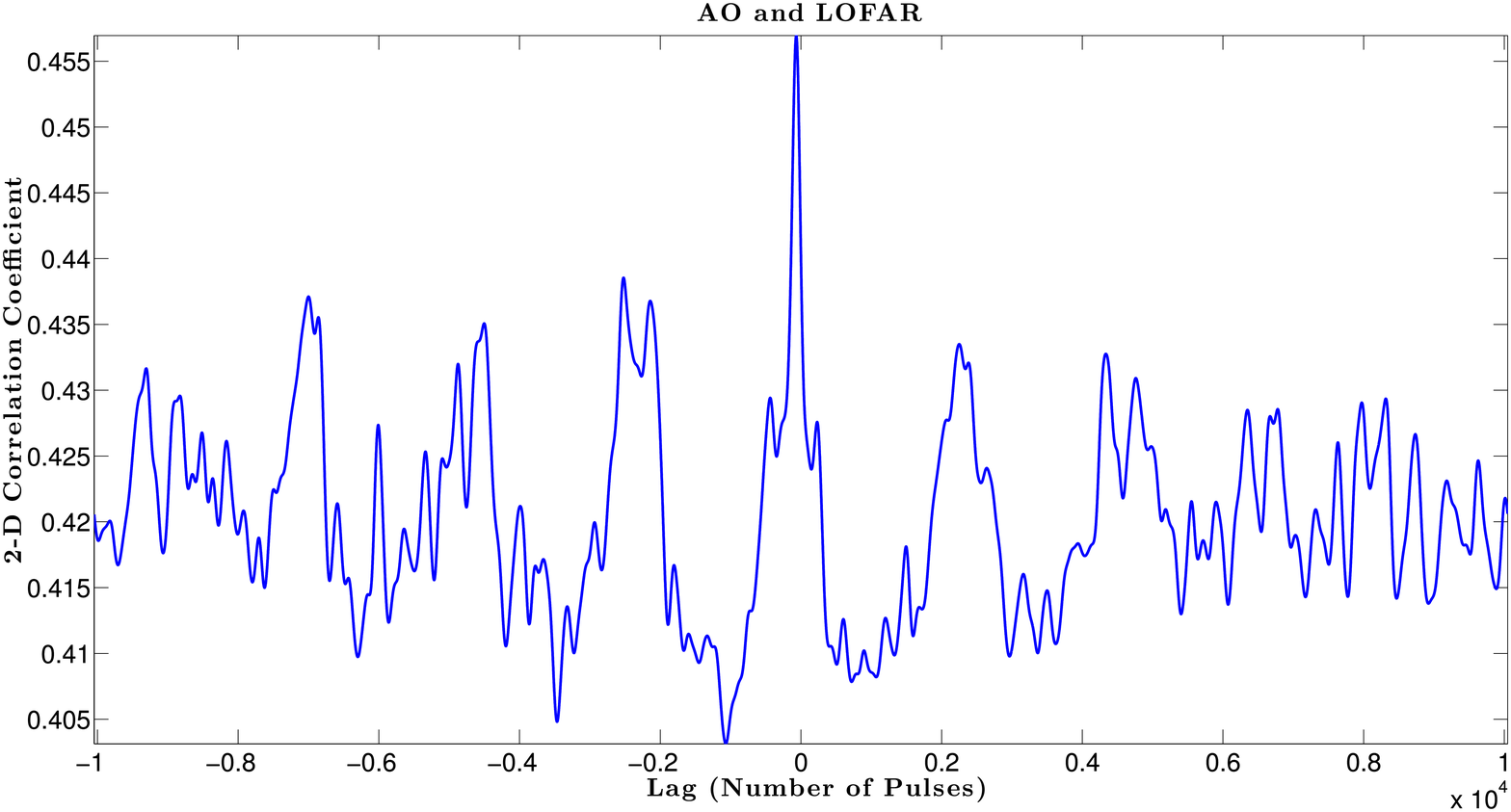,width=150mm}}


\caption{Cross-correlation function (CCF) plot of \textbf{Top Panel}: 820/150 MHz and \textbf{Bottom Panel}: 327/150 MHz. The plot clearly brings out the correlation of burst mode in 327/150 MHz and the anti-correlation in 820/150 MHz.}
\label{fig:CCF}
\end{figure*}

By doing this, we were able to obtain snippets of datasets where we
could study the bursting behaviour simultaneously at different radio frequencies. Figure~\ref{fig:LOFAR_obs} clearly shows enhanced emission to
study broadband correlations. The emission exhibits a change in its
behaviour going from 820/150~MHz observations to 327/150~MHz
observations. The direction of phase shift during bursting is different at 327 and 820~MHz when compared to their corresponding 150~MHz datasets. The correlation in 327/150 MHz and the anti-correlation
 in 820/150 MHz of the bursting is evident in Fig.~\ref{fig:CCF} which shows 
the cross-correlations between the two frequencies. Fig.~\ref{fig:CCF} 
reveals a slight offset in the maximum of the 327/150 MHz ($\sim$60 bins) and the minimum of the 
820/150 MHz ($\sim$50 bins) plot from the zero lag. Since the offsets are comparable to the 
kernel width used to smooth the datasets along the time axis, they are insignificant and most likely not intrinsic to this phenomena.

\begin{figure*} 
\centering
\begin{tabular}[ht]{|@{}l|@{}l|r@{}|}

{\mbox{\psfig{file=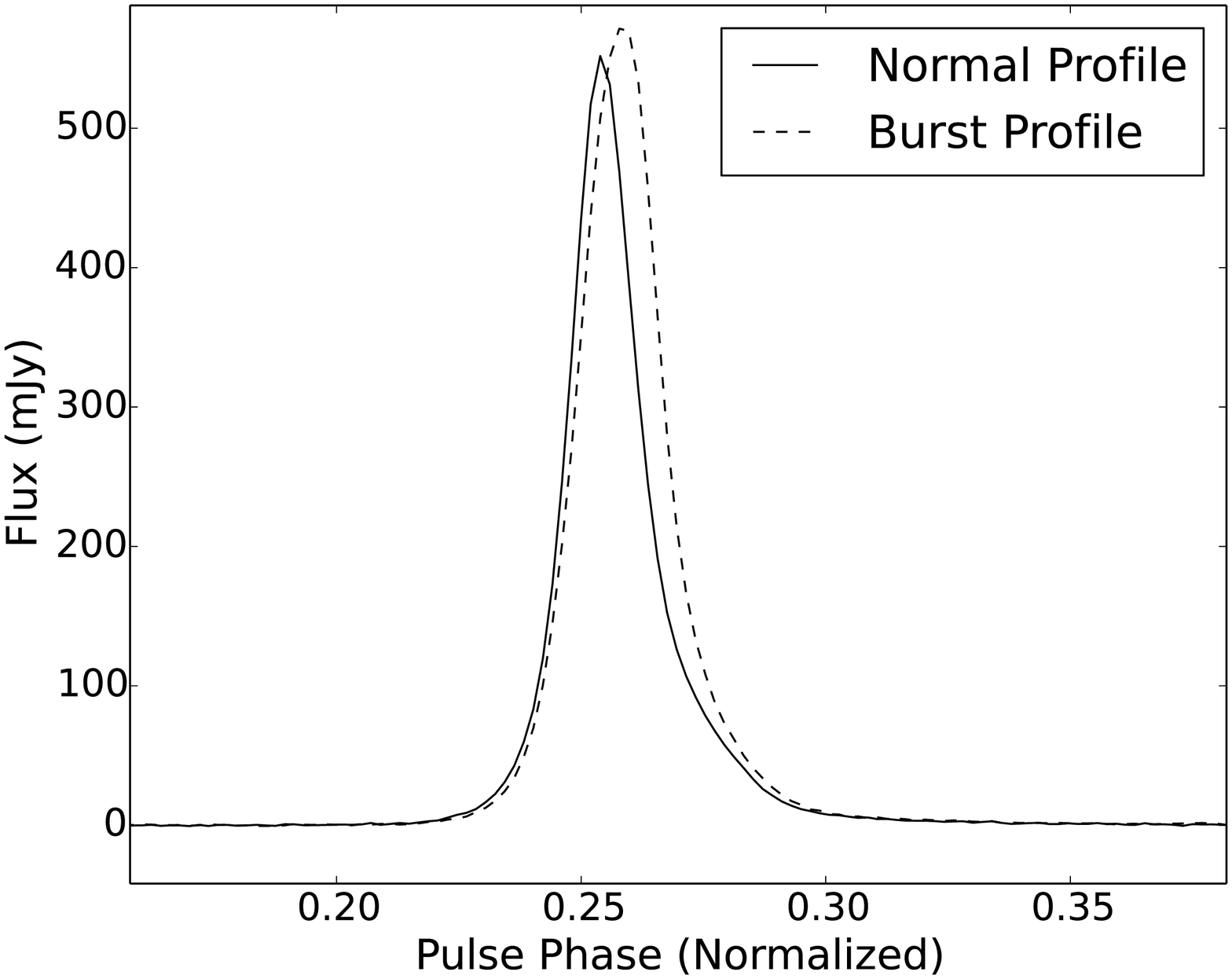,width=85mm,height=60mm}}} &{\mbox{\psfig{file=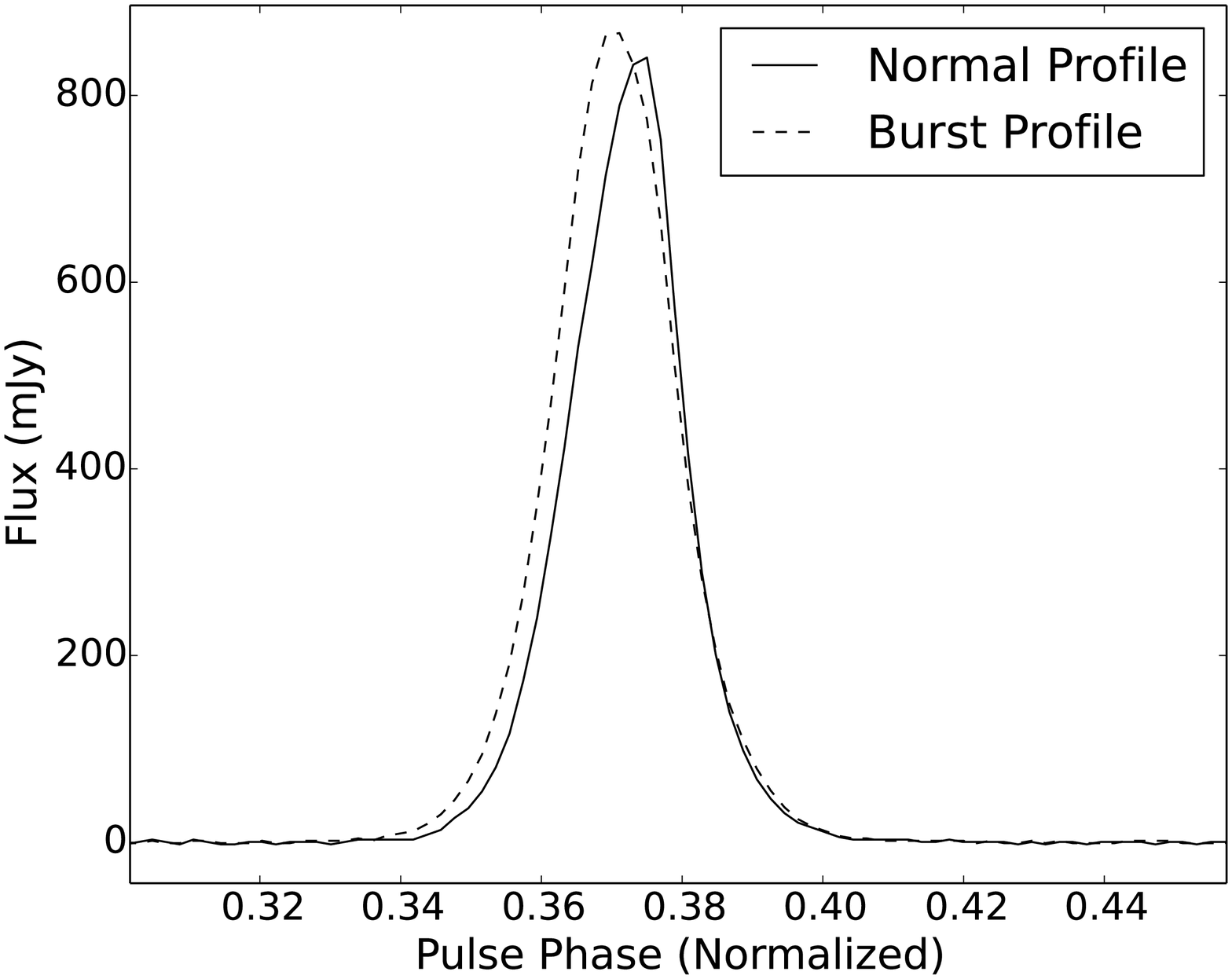 ,width=85mm,height=60mm}}}  
\\

\end{tabular}
\caption{Profiles of normal and burst mode for 327 MHz (left) and 820 MHz (right).}
\label{fig:avg_prof_norm_burst}
\end{figure*}

  We divided the time series into sections with no bursting and where
  bursting was clearly evident. The sections were selected visually
  from the smoothed data. From these sections, average profiles were
  created by summing the energies over the phase bins where the pulsed emission was seen. The
  profiles clearly bring out the difference in the behaviour of
  bursting at both frequencies. The profile for the bursting pulses is
  phase shifted in pulse longitude at the two frequencies as reported by~\cite{Se14} although there is a difference in the direction of the
  phase shift. At 327 MHz, the phase shift occurs towards the trailing
  edge of the profile while it occurs at the leading edge of the
  profile at 820 MHz. This is illustrated in Fig.~\ref{fig:avg_prof_norm_burst}.

\subsection{Spectral turnover}

As we had multi-frequency data, we investigated the spectral behaviour
of this pulsar. Using the modified radiometer equation (Lorimer \& Kramer~2012), the 
flux density
\begin{equation}
S = \frac{\beta T_{\rm sys} A}{G N_{\rm bin} \sigma_{\rm off} \sqrt{\Delta \nu N_p t_{\rm obs}}},
\label{eq:flux}
\end{equation}
where $A$ is the area under
the pulse, $N_{\rm bin}$ is the total number of phase bins in the profile, $N_p$ is the number of polarizations, $t_{\rm obs}$ is the total integration time of each phase 
bin of the pulse profile, $\beta$ is the correction factor for digitization, $\sigma_{\rm off}$ is the rms of the noise in the pulsar timeseries and the rest of the 
parameters are as given in Table~\ref{tab:config}.
 
Estimating the flux of the LOFAR observations was not a straightforward process. For FR606, which consists of antennas without moving parts, the temperature is 
strongly dependent on frequency, while the gain depends on frequency and on 
source position on the sky (elevation and azimuth). To calibrate the observed 
flux, we have used software described in detail by~\cite{Ko15}. The software 
produces flux density scaling factor using Eq~\ref{eq:flux} for each pulse and frequency channel. In order to properly estimate $G$ and $T_{sys}$, the software 
makes use of the Hamaker beam model~\citep{Ha06} and mscorpol\footnote{\url{https://github.com/2baOrNot2ba/mscorpol}} by Tobia Carozzi to calculate Jones 
matrices of the HBA antenna response for a given frequency and sky coordinates. The HBA antenna temperature, $T_{A}$, is derived from CasA measurements done by~\cite{Wi11}. The background sky temperature, $T_{sky}$, is taken from a sky map at 408~MHz by~\cite{Ha82}, scaled to the HBA frequency as $\nu^{-2.55}$~\citep{La87}. The error on the flux density calculation is 50$\%$ and can be attributed to e.g. error on system parameters, beam jitter due to the propagation in the ionosphere or strong sources contributing through the side lobes of the beam. Detailed discussion on error calculation is provided by~\cite{Bi15}.

 To obtain reliable flux estimates, it was important to consider all the
 biases that are introduced in this analysis due to the interstellar
 medium. The
first effect we considered is interstellar scattering. The profiles in Fig.~\ref{fig:Avg_prof} suggest that at no frequency is the scattering tail a significant 
fraction of the pulse period. Therefore, scattering does not significantly alter any flux 
estimates; hence we decided to not compensate for scattering in the analysis of this paper. Yet, a 
detailed treatment of the scattering will be discussed in another upcoming paper. Pulsar flux is also modulated by the free electron content along the
 line of sight. These modulations occur due to refractive interstellar
 scintillation \citep{Gu94} which has timescale of the order of days and
 diffractive interstellar scintillation \citep{Le11} which can be of the
 order of minutes. Refractive scintillation will have little effect on
 our analysis as the timescales are much larger than the average observation 
length. However, since we were comparing our fluxes with those from~\cite{Lo95} and since those fluxes were measured twenty years ago, we decided to check whether refractive scintillations would affect our spectrum. We calculated the 
flux of the pulsar at 1400~MHz from an observation done few years ago for~\citep{Se14} and found that flux to match with the Lorimer et al. (1995) flux indicating that refractive scintillation may not dramatically modulate the flux. To check if the diffractive scintillations were biasing our flux estimates, we used the
 NE2001 electron distribution model \citep{Co02}  to
 calculate the diffractive scintillation timescales at 150, 327 and
 820 MHz. The values we obtained were 46, 65 and 100~s,
 respectively, which were small compared to the total integration
 time and the timescale of intrinsic variation of pulsar flux. Hence, we could use the pulse profiles as they were for further
 analysis.

\begin{table}
\begin{tabular}{lc lc}
\hline
Frequency &Flux \\
 (MHz)    & (mJy) \\
\hline
150& 88$\pm 44$\\
327& 13.8$\pm 0.8$ \\
408& \textbf{29$\pm 1$} \\
820& 16.1$\pm 1.7$ \\
1408& \textbf{2.2$\pm 1$}\\
\hline
\end{tabular}
\caption{Flux densities calculated at different frequencies for PSR~B0611+22. The values in bold are taken from \protect\cite{Lo95}.
\label{tab:flux}}
\end{table}

\begin{figure}
\centering
\psfig{file=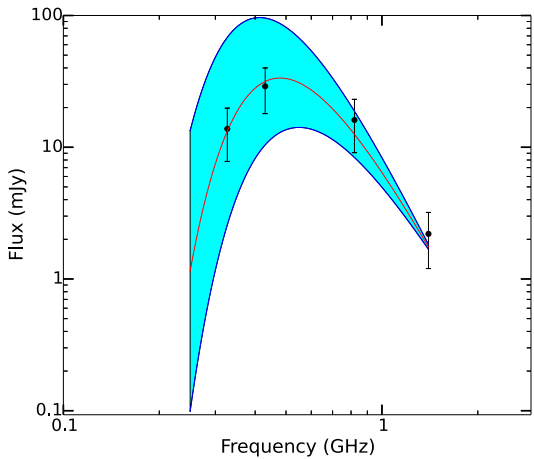,width=80mm}
\caption{Modeled Spectrum for PSR B0611+22 using a free-free absorption model. The 150~MHz flux has been excluded from the analysis. The red curve is the best fit curve. The shaded region is enclosed within $\pm$~1~$\sigma$ curves. The $\pm$~1~$\sigma$ limits are derived from the fit. The electron temperature was set to 5000~K~\citep{Ra15}. The derived emission measure was roughly 5$\times$10$^{5}$~cm$^{-6}$~pc. The reduced $\chi^{2}$ on the fit was 0.26.
\label{fig:Spec}}
\end{figure}

  The flux densities at various frequencies are listed in Table
  \ref{tab:flux}. The values suggest that the pulsar spectrum does not follow the standard power law model. To make sure that the 327 MHz flux was not lower due to refractive scintillations, we calculated the flux of another observation at the same frequency 
which was used in~\cite{Se14}. The calculated flux value was comparable to 
the new observations. In this analysis, we realized that the peak flux in our calculations did not match with peak fluxes in~\cite{Se14}. After further 
cross-checks and verification, we conclude that the flux estimates 
in~\cite{Se14} are off by a factor of 16 as in their 
calculations, the time per bin ($t_{\rm obs}$ in Eq~\ref{eq:flux}) is incorrect.
Hence, we use the flux estimates computed here for further analysis. The turnover 
we see is in the middle of the frequencies where we see a temporal  
anti-correlation in the bursts which suggests that the physical processes 
responsible for the anti-correlation in bursts  might also be responsible for 
this turnover. The other possibility is that the observed turnover is due to absorption of radio emission by an external absorber. The dense environment around the pulsar can lead to a turnover in 
the spectrum at a higher frequency than expected. This motivated us to model the spectrum using a free-free thermal absorption model to estimate the optical 
depth and the peak frequency \citep{Le15, Ra15}. Past studies have detected 
H$\alpha$ emission \citep{By79} in the region suggesting the presence of 
ionized gas. Fig.~\ref{fig:Spec} shows the best fit model to the spectrum without the 150~MHz observation. Although the dense ISM along the line of sight makes this model very tempting, the model fails to explain the whole spectrum of the pulsar.
 
\subsection{Broadband flux density modulation}

\begin{figure}
\centerline{\psfig{file=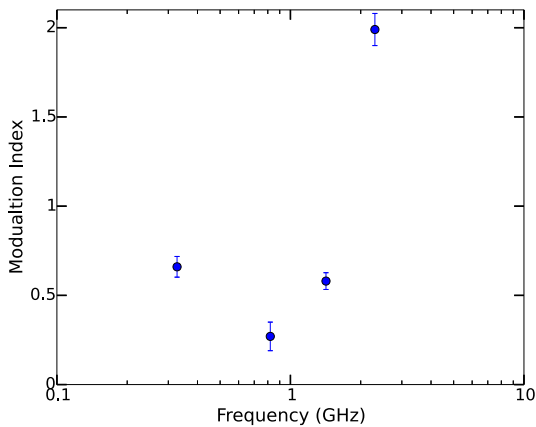,width=80mm}}
\caption{Modulation index $m_{\rm int}$ as a function of frequency for 
PSR~B0611+22. The modulation index at 2~GHz was obtained from archival unpublished data from the Green Bank Telescope.}
\label{fig:mod_ind}
\end{figure}

To quantify the variation in the detected flux from the pulsar, we estimated the
modulation index at different frequencies. Modulation index is a measure of pulse to pulse intensity fluctuation. To derive the intrinsic modulation indices ($m_{\rm int}$) we used the
method described in \cite{Kr03}. First, we estimated the flux
of each of the observations. Since we did not have calibrator
observations, we used the radiometer equation \citep{Lo12} to obtain flux densities at different
frequencies. Then, we normalized the time series by
a 200~s running median to correct the pulsar
signal for any possible effects of interstellar scintillation.
Finally, the datasets were re-scaled to be consistent with
the initial average flux density. For every observation, we integrated
individual pulses to obtain the average pulse profile and calculated
its flux density (see Table~\ref{tab:flux}). After correcting the datasets
for the effects of interstellar scintillation, for each one, we calculated the
intrinsic modulation index
\begin{equation}
\label{modidzeqn}
m_{\rm int}^2 = \frac{\langle(S - \langle S \rangle)^2 \rangle}{\langle S \rangle^2},
\end{equation}
where, $\langle S \rangle$ is the mean flux density and $S$ is the measured flux
density of a single pulse. We compared our results with results for other pulsars
\citep{Ba80, Kr03}. It is clearly seen that
$m_{\rm int}$ decreases with frequency until a certain ``cut off'' frequency and rises again
 (see Fig.~\ref{fig:mod_ind}). Our results are consistent with other
pulsars where a similar trend is observed \citep{Ma13}.

\subsection{X-ray flux upper limits}

We observed PSR~B0611+22 for 33~ks using {\it XMM Newton}. We observed the pulsar using both PN CCD and MOS camera mounted on the telescope.
We did not detect the pulsar but obtained an upper limit
on the X-ray flux using the method described in~\cite{Lo12}. We obtained
an upper limit on the count rate of $\sim5\times$10$^{-4}$ counts s$^{-1}$
at 99\% confidence level. Then, assuming a non-thermal emission from the pulsar
with a photon index $\Gamma \approx 2$~\citep{Pa09} and using the publicly available software
\textbf{WEBPIMMS}\footnote{\url{https://heasarc.gsfc.nasa.gov/cgi-bin/Tools/w3pimms/w3pimms.pl}},
we obtained a limit on the unabsorbed X-ray flux, $F_{0.3-8\,{\rm keV}}^{\rm
unabs}$of$\sim$$2.7 \times 10^{-15}$~ergs~cm$^{-2}$~s$^{-1}$ assuming a neutral
hydrogen column density $N_{H}$ of $3.1 \times 10^{21}$~cm$^{-2}$, estimated using 10\%
ionization fraction. The obtained upper limit is less than the flux reported for PSR B0943+10~\citep{He13} using the same non-thermal model for fitting. From this flux, we estimated an upper limit of
X-ray luminosity, $L_X = 4\pi d^2 F_X^{\rm unabs}$ to be $< 2.7 \times 10^{30}$~ergs~s$^{-1}$. 
From there, we were able to obtain an
upper bound on the X-ray efficiency $\eta_{\rm 0.3-8\,keV} 10^{-5}$ by assuming an $\dot E$=$10^{34}$~ergs~$\rm s^{-1}$. 
This upper limit has an errorbar of roughly 50$\%$ given the uncertainties in the distance to the 
source.

\section{Discussion}

\subsection{Quasi-stable magnetosphere?}

A detailed understanding of the physics of pulsar magnetospheres has 
presented a major challenge to astronomers over the past four
decades. Recently, new observations have shed some light on emission
physics.~\cite{Ly10}
reported a correlation between the spin-down rate
and pulse profile changes in a sample of mode changing pulsars. They concluded that the magnetosphere switched between
multiple quasi-stable states of emission. This hypothesis was bolstered by recent radio and X-ray observations of PSR B0943+10, which showed that the thermal X-ray emission was anti-correlated with the radio emission
\citep{He13}.

Based on Rankin's model of radio emission
from neutron stars~\citep{Ra83}, PSR~B0611+22 is a single core
component pulsar. 
Later, \cite{Ly88} and \cite{Jo08} classified this pulsar as a partial 
cone based on the polarization position angle sweep across the profile. 
Partial cone pulsars are known to exhibit broadband phenomena like nulling and mode changing~\citep{Bh07,Ga14}.
We observed that in PSR B0611+22,
during the burst, the phase of the pulse shifts slightly, which agrees with the results in~\cite{Se14}. In
the 820~MHz observation, the bursting occurred when the pulse phase
shifted towards the leading edge. On the other hand, the bursting occurred
when the pulse phase shifted towards the trailing edge in the 327~MHz
observations. This opposing behaviour at different frequencies leads
to an anti-correlation when compared to the simultaneous 150~MHz data.
 Though PSR~B0611+22 seems to be a mode 
changing pulsar, the frequency dependence of the mode change is not explained 
by current mode changing models and observations~\citep{Ra83}.~\cite{Ke00} claims that phase offset in the 
bursting profile occurs due to the periodic existence of a conal component along 
with the core component. Though the model is able to explain the slight 
increase in the width of the pulse profile in bursting mode, it fails to 
account for the different direction of phase offset at different frequencies. 
The observations presented here suggest that the spectral indices of the two components are different. Therefore, if the components preserve their phase,  this manifests 
itself as an anti-correlation in the two simultaneous datasets We intend to do 
follow-up observations with a wider frequency coverage before making any claims on the model of emission of this pulsar. 

 At 327 and 820~MHz, we observed quasi-periodic bursting behaviour. 
Though the period can be derived from the peak in the power 
spectrum, the lack of harmonics suggests that we need to sample more bursts to better characterize the periodicity. This 
also puts forth a question of whether similar pulsars like PSR J1752+2359~\citep{Ga14} and PSR J1939+2213 exhibit such
behaviour.

The non-detection of the pulsar in the X-ray waveband contradicts the 
predictions made in~\cite{Se14}, based on assuming the 1$\%$ X-ray efficiency 
of PSR~B0943+10. However, it is consistent with average efficiency of $8\times10^{-5}$ found in~\cite{Ja11} for old (age $>$ 17 kyr) pulsars and their nebulae.
This shows that the pulsar is a weak X-ray emitter. The assumption of efficiency of 1$\%$ in~\cite{Se14} is based on PSR~B0943+10 which has different spin 
down properties compared to PSR~B0611+22. Hence, it is not surprising that the 
efficiency is dramatically different.

\subsection {Flux Density Spectrum}

From our flux density estimates, we were able to conclude that the spectrum of PSR B0611+22 does not exhibit single power law behaviour. This could be due to free-free thermal absorption along the line of sight. Recently,
\cite{Ki11} and \cite{De14} have shown that these so called giga hertz peaked (GPS) spectra pulsars have peculiar
environments in close vicinity of the pulsar or potential absorbers lying along 
the line of sight. This clearly suggests a turnover might occur at a frequency 
higher than 100~MHz. Recent modeling and simulations
\citep{Le15, Ra15} suggest that free-free absorption in high density plasma surrounding supernova remnants, pulsar wind nebulae or 
dense, cold, partially ionized gas is responsible for high frequency turnovers.
 Using this idea, we modeled the spectrum with a free-free thermal absorption 
model. The main caveat in this result is that the model cannot explain the 
whole spectrum of the pulsar because of the higher flux value of 150~MHz 
observation even though free-free thermal absorption seems like a promising 
explanation to the non-power law behaviour given 
the dense environment along the line of sight toward the pulsar.

Since the model cannot explain the whole spectrum, external absorption scenario seems unlikely. The other possibility we considered is that the 
irregular spectral behaviour could be caused due to the intrinsic variations 
in the pulsar itself. The different spectral indices of the two components within the partial cone could appear as a turnover in the flux density spectrum. Future observations at frequencies higher than 820 MHz will shed some light on 
this phenomenon. Also, it will be important to verify the flux densities at LOFAR frequencies for future studies. 

We also calculated the modulation indices at various
frequencies. It can be seen (Fig.~\ref{fig:mod_ind}) that there is a turnover at $\sim1$~GHz. This result agrees with what is seen in other pulsars
\citep{Ma00}. The turnover in the modulation indices is mainly caused by
decreasing average pulsar energy. Therefore the number of so-called pseudo nulls
(no detection due to inadequate receiver sensitivity) increases with frequency.

\section{Conclusions}

We have carried out a detailed analysis of simultaneous radio and X-ray
observations of pulsar PSR~B0611+22. The multi-frequency data reveal
a wealth of information about the emission characteristics of this
pulsar. The bursting behaviour varied across the radio band with a 
quasi-periodic characteristic at all frequencies. The 327/150 MHz and 820/150 MHz simultaneous 
observations show an anti-correlation in bursting. We leave modeling this unusual behaviour to a later paper.  
Future polarimetric studies of both modes will help in discerning the emission physics of this pulsar. Moreover, we obtained a flux density spectrum from
the radio observations of this pulsar. The spectrum shows a turnover at higher frequencies. We considered free-free thermal absorption by the surrounding ISM as a possible explanation 
but such model cannot explain the flux density at 150~MHz. From the X-ray 
non-detection, we obtained an upper bound on the X-ray luminosity and X-ray efficiency
of the pulsar. The X-ray non-detection shows that the X-ray efficiency is low and consistent with X-ray efficiencies measured for other similarly aged pulsars.

\section*{Acknowledgements}
We thank our referee for useful comments that greatly improved the manuscript.
This work was supported by a NASA XMM-Newton grant (award no. 74279). We thank West Virginia University for its financial support of GBT operations, which enabled some of the observations for this project.
We thank the staff of LOFAR station, Nan\c{c}ay, France for the observations.
The financial assistance of the South African SKA project (SKA SA) towards this research is hereby acknowledged. Opinions expressed and conclusions arrived at are those of the authors and are not necessarily to be attributed to the SKA SA.

\bibliography{B0611ref}{}
\bibliographystyle{mn2e}

\end{document}